\documentclass[aps,12pt,pra,superscriptaddress,preprint,a4paper]{revtex4}
\usepackage{amsmath}
\usepackage{subfig}
\usepackage{array}
\usepackage{amsfonts}
\usepackage{epsf}
\usepackage{epsfig}
\usepackage{amssymb}
\usepackage{bbm}
\usepackage{xcolor}
\numberwithin{equation}{section}

\begin{document}
\begin{center}
\end{center}
\title{Time independent fractional Schr\"{o}dinger equation for generalized Mie-type potential in higher dimension framed with Jumarie type fractional derivative}
\author{\small Tapas Das }
\email[E-mail: ]{tapasd20@gmail.com}\affiliation{Kodalia Prasanna Banga High School (H.S), South 24 Parganas 700146, India}
\author{\small Uttam Ghosh }
\email[E-mail: ]{uttam_math@yahoo.co.in}
\author{\small Susmita Sarkar}
\email[E-mail: ]{susmita62@yahoo.co.in}\affiliation{Department of Applied Mathematics, University of Calcutta, Kolkata, India}
\author{\small Shantanu Das}
\email[E-mail: ]{shantanu@barc.gov.in}\affiliation{Reactor Control System Design Section (E \& I Group) BARC Mumbai India}
\begin{abstract}
In this paper we obtain approximate bound state solutions of $N$-dimensional fractional time independent Schr\"{o}dinger equation for generalised Mie-type potential, namely $V(r^{\alpha})=\frac{A}{r^{2\alpha}}+\frac{B}{r^{\alpha}}+C$. Here $\alpha(0<\alpha<1)$ acts like a fractional parameter for the space variable $r$. When $\alpha=1$ the potential converts into the original form of Mie-type of potential that is generally studied in molecular and chemical physics. The entire study is composed with Jumarie type fractional derivative approach. The solution is expressed via Mittag-Leffler function and fractionally defined confluent hypergeometric function. To ensure the validity of the present work, obtained results are verified with the previous works for different potential parameter configurations, specially for $\alpha=1$. At the end, few numerical calculations for energy eigenvalue and bound states eigenfunctions are furnished for a typical diatomic molecule.  
\end{abstract}
\maketitle
\section{I\lowercase{ntroduction}}
Although the classical calculus provides quite promising tool to model and explain different dynamical phenomenon, still there are uncountable complex systems in nature which demand the generalization of classical calculus. For example, the non-linear oscillation of earthquake [1], the fluid dynamics traffic model [2], the transport of chemical contaminant through porous rock [3], the dynamics of visco-elastic materials [4], and more in biology, continuum and statistical mechanics, anomalous transport, magneto-thermoelastic diffusion [5-8] are only possible to model elegantly, at least from mathematical point of view, if concern differential equations are developed with arbitrary order derivatives (commonly known as fractional order derivatives).\\
Addition to the above list, quantum mechanics is also one of the most celebrated area of study where fractional derivatives are employed for remodelling the Schr\"{o}dinger equation [9]. Basically being a non-relativistic equation, Schr\"{o}dinger equation is an essential tool for the study of atoms, nuclei and molecules and their spectral behaviours through proper potential function which describes the nature of bonding of the quantum particles. One may ask `why we will remodel Schr\"{o}dinger equation via fractional derivatives?' `Does this change will describe any new phenomena?' Obviously there is a physical reason behind this. There are lots of similarities between standard diffusion equation and basic Schr\"{o}dinger equation. Now when non-Brownian path are taken in the path integral approach, then resulting diffusion equation emerges with fractional derivatives which properly describes many complex diffusion phenomenon [10]. As Schr\"{o}dinger equation is an outcome of Feynman path integral formalism one may expect same type of outcome for Schr\"{o}dinger equation over the non-Brownian path. Based on this theoretical view fractional Schr\"{o}dinger equation has been developed by Laskin [11]. After then Guo and Xu [12], Dong and Xu [13] studied the space fractional Schr\"{o}dinger equation with few specific potential models. More recent works on the one dimensional fractional Schr\"{o}dinger equation can be found in reference [14-17].\\
Motivated by these studies and others in this paper we find the approximate bound state solution of time independent fractional Schr\"{o}dinger equation for generalized Mie-type potentials in $N$-dimensions. The idea of higher dimension is not fancy in physics. String theory, the only self-consistent quantum theory of gravity, needs more extra spatial dimensions over the usual three dimensions to support the theoretical structure. We do not feel the addition dimensions neither they have been discovered from any high energy particle physics experiments but we can not say that they do not exit. Perhaps we could be restrained to 3D world, which is in fact of part of a more complicated mutidimensional universe possibly up to 10 or more spacial dimensions [18-19]. The advantage of higher dimensional study is, it provides a general treatment of the model in such a manner that one can obtain required results in lower dimensions just dialling appropriate $N$. The Mie-type potentials [20] originally defined as
\begin{eqnarray*}
V(r)=D_0\left[ \frac{a_0}{b_0-a_0}\left( \frac{r_0}{r}\right)^{b_0}-\frac{b_0}{b_0-a_0}\left( \frac{r_0}{r}\right)^{a_0}\right] \,,
\end{eqnarray*} 
where $D_0$ is the interaction energy between two atoms in a molecular system at equilibrium distance $r=r_0$. Here $a_0$ and $b_0$ are restricted to take integer values. In this paper we will overrule the restriction for $a_0$ and $b_0$ and proposed a more general form of the potential by adding a constant term $C$. If $b_0=2\alpha$, $a_0=\alpha$ $(0<\alpha<1)$ then the potential gets a nicer form as
\begin{eqnarray}
V(r^{\alpha})=\frac{A}{r^{2\alpha}}+\frac{B}{r^{\alpha}}+C\,,
\end{eqnarray} 
where $A=D_0r_0^{2\alpha}$ and $B=-2D_0r_0^{\alpha}$. This form is customary and flexible in terms mathematical point of view. When $\alpha=1$ with $A=-D_0r_0^2, B=2D_0r_0 $ and $C=-D_0$, we have the modified Kartzer potential and similarly  for $A=D_0r_0^2, B=-2D_0r_0 $ and $C=0$ we have Kartzer-Fues potential [21]. \\
To make this present work self contained it is organized as follows: In the next section we briefly outline Jumarie type fractional derivative and Laplace transform. In section 3 we construct the $N$-dimensional Schr\"{o}dinger equation via fractional Laplacian operator in hypersherical coordinate. Bound state spectrum for the Mie type potential is obtained in section 4. Section 5 is devoted for the discussion where theoretical as well as numerical results are discussed with few eigenfunctions plotting. Finally the conclusion of the work appears in the section 6.
\section{O\lowercase{utline of fractional derivative and} L\lowercase{aplace transform}}
\subsection{Fractional order derivative of Jumarie type}
Jumarie [22] defined the fractional order derivative by modifying the left-Riemann-Liouville (RL) fractional derivative in the following form for a continuous function $f(x)$ (but not necessarily differentiable) in the interval $a$ to $x$, with $f(x)=0$ for $x<a$
\begin{eqnarray}
\,_a^{J}D_{x}^{\alpha}[f(x)]=f^{(\alpha)}(x)=\begin{cases} \frac{1}{\Gamma(-\alpha)}\int_a^{x}(x-\xi)^{-\alpha-1}f(\xi)d\xi,&   \alpha<0,\\
\frac{1}{\Gamma(-\alpha)}\frac{d}{dx}\int_a^{x}(x-\xi)^{-\alpha}(f(\xi)-f(a))d\xi,&  0<\alpha<1,\\
(f^{(\alpha-n)}(x))^{(n)},&  n \leq\alpha<n+1.
\end{cases}
\end{eqnarray}
It is customary to take the start point of the interval as $a=0$ and use the symbol $\,_0^{J}D_{x}^{\alpha}[f(x)]$ for $f^{(\alpha)}(x)$. Here from in the rest of the paper we will always denote the fractional derivative $f^{(\alpha)}(x)\equiv\frac{d^\alpha f(x)}{dx^\alpha}$ as $\,_0^{J}D_{x}^{\alpha}[f(x)]$ with Jumarie sense. In the above definition, the first expression is just the Riemann-Liouville fractional integration, the second expression is known as modified Riemann-Liouville derivative of order $0<\alpha<1$ because of the involvement of $f(a=0)$. The third line definition is for the range $n\leq\alpha<n+1$. Apart from the integral type of definition we can also express fractional derivative via fractional difference. Let $f:\Re\rightarrow\Re$, denotes a continuous (but not necessarily differentiable) function such that $x\rightarrow f(x)$ for all $\in\Re$. If $h>0$ denotes a constant discretization span with forward operator $FW(h)f(x)=f(x+h)$; then the right hand fractional difference of $f(x)$ of order $\alpha$ ($0<\alpha<1$) is defined by the expression [23]
\begin{eqnarray}
\bigtriangleup^{\alpha}f(x)=(FW(h)-1)^{\alpha}f(x)=\sum_{i=0}^{\infty}(-1)^i{{\alpha}\choose{i}}f[x+(\alpha-i)h]\,,
\end{eqnarray}
where generalized binomial coefficients $\frac{\Gamma(\alpha-i)}{\Gamma(-\alpha)\Gamma(i+1)}={{i-\alpha-1}\choose{i}}=(-1)^i{{\alpha}\choose{i}}$. These equalities being readily established from the definition of a binomial coefficient and generalization of factorials with Gamma function $\,^nC_r={{n}\choose{r}}=\frac{n!}{r!(n-r)!}$. 
Then the Jumarie fractional derivative is defined as
\begin{eqnarray}
f^{(\alpha)}(x)=\lim_{h\downarrow 0}\frac{\bigtriangleup^{\alpha}[f(x)-f(0)]}{h^{\alpha}}=\frac{d^{\alpha}f(x)}{dx^{\alpha}}\,.
\end{eqnarray}
This definition is close to the standard definition of derivatives for beginner's study. Following this definition it is clear that the $\alpha$-th derivative of a constant for $0<\alpha<1$ is zero. Few results for Jumarie type derivative are listed below depending on the characteristics of given function $(f[u(x)])$ [24]
\begin{subequations}
\begin{align}
\,_0^{J}D_{x}^{\alpha}(f[u(x)])&=f_{u}^{(\alpha)}(u)(u_x^{'})^{\alpha}\,,\\
\,_0^{J}D_{x}^{\alpha}(f[u(x)])&=(f/u)^{1-\alpha}(f^{'}_u(u))^{\alpha}u^{\alpha}(x)\,,\\
\,_0^{J}D_{x}^{\alpha}(f[u(x)])&=(1-\alpha)!u^{\alpha-1}f_u^{(\alpha)}(u)u^{\alpha}(x)\,,\\
\,_0^{J}D_{x}^{\alpha}(x^{\beta})&=\frac{\Gamma(1+\beta)}{\Gamma(1+\beta-\alpha)}x^{\beta-\alpha}\,. 
\end{align}
\end{subequations}
In fractional calculus solution of any linear fractional differential equation, composed with Jumarie derivative, can be easily obtained in terms of Mittag-Leffler function of one parameter [25] which is defined as
\begin{eqnarray}
E_{\alpha}(z)=\sum_{\kappa=0}^{\infty}\frac{z^\kappa}{\Gamma(\alpha \kappa+1)}\,,\,\, (\alpha>0)\,.
\end{eqnarray} 
or more general form [26] $E_{\alpha,\beta}(z)=\sum_{\kappa=0}^{\infty}\frac{z^\kappa}{\Gamma(\alpha \kappa+\beta)}$. Clearly $E_{\alpha,1}(z)=E_{\alpha}(z)$ and $E_{1,1}(z)=E_1(z)=e^z$. We provide few derivative rules [27-28] associated with the Mittag-Leffler function and its trigonometric counterparts.
\begin{subequations}
\begin{align}
\,_0^{J}D_{x}^{\alpha}[E_{\alpha}(ax^{\alpha})]&=aE_{\alpha}(ax^{\alpha})\,,\\
\,_0^{J}D_{x}^{\beta}[E_{\alpha}(ax^{\alpha})]&=x^{\alpha-\beta}E_{\alpha,\alpha-\beta+1}(x^{\alpha})\,,\\
\,_0^{J}D_{x}^{\alpha}[cos_{\alpha}(ax^{\alpha})]&=-asin_{\alpha}(ax^{\alpha})\,,\\
\,_0^{J}D_{x}^{\alpha}[sin_{\alpha}(ax^{\alpha})]&=acos_{\alpha}(ax^{\alpha})\,,
\end{align}
\end{subequations}
where one parameter fractional sine and cosine function are defined as follows [29]\\
$cos_{\alpha}(x^{\alpha})=\sum_{\kappa=0}^{\infty}(-1)^\kappa \frac{x^{2\kappa\alpha}}{\Gamma(1+2\alpha \kappa)}$ and $sin_{\alpha}(x^{\alpha})=\sum_{\kappa=0}^{\infty}(-1)^\kappa \frac{x^{(2\kappa+1)\alpha}}{\Gamma(1+(2\kappa+1)\alpha )}$ with \\$E_{\alpha}(ix^{\alpha})=cos_{\alpha}(x^{\alpha})+isin_{\alpha}(x^{\alpha})$.  
\subsection{Laplace transformation of fractional differ-integrals}
In general Laplace transform $F(s)$ or $\mathcal{L}$ of a function $f(x)$ is defined as [30]
\begin{eqnarray}
F(s)=\mathcal{L}\left\{f(x)\right\}=\int_0^\infty e^{-sx}f(x)dx\,.
\end{eqnarray}
If there is some constant $\sigma\in\Re$ such that $|e^{-\sigma x}f(x)|\leq\mathcal{M}$ for sufficiently large $x$, the above definition will exist for Re $[s]>\sigma$. The following are the well known derivative properties of Laplace transform when $n$ is an integer.
\begin{subequations}
\begin{align}
\mathcal{L}\left\{f^{(n)}(x)\right\}&=s^nF(s)-\sum_{k=0}^{n-1}s^{n-1-k}{f^{(k)}(0)}\,,\\
\mathcal{L}\left\{x^{n}f(x)\right\}&=(-1)^{n}F^{(n)}(s)\,,
\end{align}
\end{subequations}
where the superscript $(n)$ denotes the $n$-th derivative with respect to $x$ for $f^{(n)}{(x)}$, and with respect to $s$ for $F^{(n)}(s)$. Now if $n$ becomes non integer, say $\alpha$, then above two rules are generalized as [31]
\begin{subequations}
\begin{align}
\mathcal{L}\left\{\frac{d^{\alpha}f(x)}{dx^{\alpha}}\right\}&=s^\alpha F(s)-\sum_{k=0}^{n-1}s^k\frac{d^{\alpha-1-k}f(x)}{dx^{\alpha-1-k}}\Bigg|_{x=0}\,,\\
\mathcal{L}\left\{x^{\alpha}f(x)\right\}&=-\tau\frac{d^\alpha F(s)}{ds^\alpha}\,,
\end{align}
\end{subequations}
where, $n$ is the largest integer such that $(n-1)<\alpha\leq n$ and $\tau=-\frac{cosec((\alpha-\delta)\pi)}{cosec(-\delta \pi)}$ with $-1<\delta<0$ (see Appendix-2). The sum in the Eq.(2.9a) is zero when $\alpha\leq 0$.  The Eq.(2.9b) shows that $\mathcal{L}\left\{x^{\alpha}f(x)\right\}\neq \pm\frac{d^\alpha F(s)}{ds^\alpha}$ but when $\alpha=1$, (that makes $\tau=1$) this is consistent with Eq.(2.8b) if we take $n=1$. Choosing the initial condition $f(0)=0$ (frequently appears in quantum mechanical problems) it is easy to have $\mathcal{L}\left\{\frac{d^{\alpha}f(x)}{dx^{\alpha}}\right\}=s^\alpha F(s)$. Under this circumstance one can generate $\mathcal{L}\left\{x^\alpha\frac{d^{\beta}f(x)}{dx^{\beta}}\right\}$ as follows
\begin{align}
\mathcal{L}\left\{x^\alpha\frac{d^{\beta}f(x)}{dx^{\beta}}\right\}&=-\tau\frac{d^\alpha}{ds^\alpha}\mathcal{L}\left\{\frac{d^{\beta}f(x)}{dx^{\beta}}\right\}\,,\nonumber\\
&=-\tau\frac{d^\alpha}{ds^\alpha}[s^\beta F(s)]\nonumber \\
&=-\tau\Big[F(s)\frac{\Gamma(1+\beta)}{\Gamma(1+\beta-\alpha)}s^{\beta-\alpha}+s^\beta\frac{d^\alpha F(s)}{ds^\alpha}\Big]\,,
\end{align}
where we have used the rule $(uv)^\alpha=u^{(\alpha)}v+v^{(\alpha)}u$ in Jumarie sense.  
Beside the above, the following operational formulae are well established [25,31]
\begin{subequations}
\begin{align}
\mathcal{L}\left\{x^\alpha\right\}&=\frac{\Gamma(1+\alpha)}{s^\alpha}\,,\\
\mathcal{L}\left\{E_\alpha(ax^\alpha)\right\}&=\frac{s^{\alpha-1}}{s^\alpha-a}\,,\\
\mathcal{L}\left\{E_\alpha^{(k)}(-x^\alpha)\right\}&=\frac{s^{\alpha+k-1}}{s^\alpha+1}\,,\\
\mathcal{L}\left\{x^{\alpha k+\beta-1}E_{\alpha,\beta}^{(k)}(\pm ax^\alpha)\right\}&=\frac{k!s^{\alpha-\beta}}{(s^{\alpha}\mp a)^{k+1}}\,,
\end{align}
\end{subequations}
 
\section{C\lowercase{onstruction of fractional} S\lowercase{chr\"{o}dinger equation in} $N$-\lowercase{dimension}}
In this subsection we will define the Cartesian coordinates $x_i$ in $N$-dimensional space as
\begin{align}
x_1 &= r^{\alpha} cos_{\alpha}\theta_1^{\alpha}sin_{\alpha}\theta_2^{\alpha} \cdots sin_{\alpha}\phi^{\alpha}\nonumber\,,\\
x_2 &= r^{\alpha} sin_{\alpha}\theta_1^{\alpha}sin_{\alpha}\theta_2^{\alpha} \cdots sin_{\alpha}\phi^{\alpha}\nonumber\,,\\
x_b &= r^{\alpha} cos_{\alpha}\theta_{b-1}^{\alpha}sin_{\alpha}\theta_b^{\alpha} \cdots sin_{\alpha}\phi^{\alpha}\,,\\
x_N &= r^{\alpha} cos_{\alpha}\phi^{\alpha}\nonumber\,,
\end{align}
where $b\in[3, N-1]$ and $\phi^{\alpha}=\theta_{N-1}^{\alpha}$. Here $r^{\alpha}$ and $\theta_b^{\alpha}$ are the hyperspherical coordinates in $N$-dimensions. Clearly when $N=3$ and $\alpha=1$ the above equations convert into usual three dimensional coordinate system $(r,\theta,\varphi)$. Denoting $\theta_1=\varphi$ , $\theta_2(=\phi)=\theta$ we can have $x_1\equiv x=rsin\theta cos\varphi, x_2\equiv y=rsin\theta sin\varphi, x_3\equiv z=rcos\theta$. Proceeding further to the present hyperspherical coordinates the sum of the squares of Eqs.(3.1) provides
\begin{eqnarray}
r^{2\alpha}=\sum_{i=1}^{\infty}x_i^2\,,
\end{eqnarray}
where we have used $E_{\alpha}(\pm ix^{\alpha})=cos_{\alpha}(x^{\alpha})\pm i sin_{\alpha}(x^{\alpha})$ and $E_{\alpha}(i(x+y)^{\alpha})=E_{\alpha}(ix^{\alpha})\times E_{\alpha}(iy^{\alpha})$ with Jumarie [29] sense. Thus $r^{\alpha}$ is the fractional radius of a $N$-dimensional sphere. Now following the usual expression of Laplacian operator in curvilinear coordinate [32], the fractional Laplacian operator in terms of hyperspherical coordinates can be written as
\begin{eqnarray}
\nabla_{N}^{2\alpha}=\frac{1}{h_{\alpha}}\sum_{k=0}^{N-1}\,_0^{J}D_{\theta_k}^{\alpha}\Big(\frac{h_{\alpha}}{h_k^{2\alpha}}\,_0^{J}D_{\theta_k}^{\alpha}\Big)\,,
\end{eqnarray}  
where the symbol $\,_0^{J}D_{\theta_k}^{\alpha}=\frac{\partial^{\alpha}}{\partial\theta_k^{\alpha}}\Big(\text{or}\, \frac{d^{\alpha}}{d\theta_k^{\alpha}}\Big)$ denotes Jumarie fractional derivative operator and 
\begin{eqnarray}
r^{\alpha}=\theta_{0}^{\alpha}\,,\,\; h_{\alpha}=\prod_{k=0}^{N-1}h_k^{\alpha}\,,\,\; h_k^{2\alpha}=\sum_{i=1}^N [\,_0^{J}D_{\theta_k}^{\alpha}(x_i)]^2\,.
\end{eqnarray}
Now using the formula Eq.(2.4d), Eq.(2.6c) and Eq.(2.6d) it is not hard to find
\begin{align}
h_0^{\alpha} &= \alpha! \,,\nonumber \\
h_1^{\alpha} &= r^{\alpha}sin_{\alpha}\theta_2^{\alpha}sin_{\alpha}\theta_3^{\alpha}\cdots sin_{\alpha}\phi^{\alpha}\,,\nonumber \\
h_2^{\alpha} &= r^{\alpha}sin_{\alpha}\theta_3^{\alpha}sin_{\alpha}\theta_4^{\alpha}\cdots sin_{\alpha}\phi^{\alpha}\,,\\
h_p^{\alpha} &= r^{\alpha}sin_{\alpha}\theta_{p+1}^{\alpha}sin_{\alpha}\theta_{p+2}^{\alpha}\cdots sin_{\alpha}\phi^{\alpha}\,,\nonumber \\
h_{N-1}^{\alpha} &= r^{\alpha}\,. \nonumber 
\end{align}
So we have
\begin{eqnarray}
h_{\alpha}=\alpha!(r^{\alpha})^{N-1}sin_{\alpha}\theta_2^{\alpha}sin^{2}_{\alpha}\theta_3^{\alpha}sin^{3}_{\alpha}\theta_4^{\alpha}\cdots sin^{N-3}_{\alpha}\theta_{N-2}^{\alpha}sin^{N-2}_{\alpha}\phi^{\alpha}\,. 
\end{eqnarray}
This helps to write the Eq.(3.3) as
\begin{equation}
\nabla_{N}^{2\alpha}=\frac{1}{(\alpha!)^2}\frac{1}{(r^{\alpha})^{N-1}}\frac{\partial^{\alpha}}{\partial r^{\alpha}}\Big[(r^{\alpha})^{N-1}\frac{\partial^{\alpha}}{\partial r^{\alpha}}\Big]-\frac{\Lambda^{2\alpha}_{N-1}}{r^{2\alpha}}\,,
\end{equation} 
where $\Lambda^{2\alpha}_{N-1}$ is fractional hyperangular momentum operator. The explicit form is
\begin{eqnarray}
\Lambda^{2\alpha}_{N-1}=-\Bigg[\sum_{k=1}^{N-2}\frac{1}{sin^{2}_{\alpha}\theta^{\alpha}_{k+1}sin^{2}_{\alpha}\theta^{\alpha}_{k+2}\cdots sin^{2}_{\alpha}\phi^{\alpha}}\Big(\frac{1}{sin^{k-1}_{\alpha}\theta^{\alpha}_{k}}\frac{\partial^{\alpha}}{\partial\theta_k^{\alpha}}sin^{k-1}_{\alpha}\theta^{\alpha}_{k}\frac{\partial^{\alpha}}{\partial\theta_k^{\alpha}}\Big)+\nonumber\\ \frac{1}{sin^{N-2}_{\alpha}\phi^{\alpha}}\frac{\partial^{\alpha}}{\partial\phi^{\alpha}}\Big(sin^{N-2}_{\alpha}\phi^{\alpha}\frac{\partial^{\alpha}}{\partial\phi^{\alpha}}\Big)\Bigg]\,.
\end{eqnarray} 
Before going further here we can verify the form of usual three dimensional Laplacian operator when $\alpha=1$. Earlier we have denoted $\theta_1=\varphi$ , $\theta_2(=\phi)=\theta$. Then Eq.(3.7) gives $\nabla_3^2\equiv \nabla^2=\frac{1}{r^2}\frac{\partial}{\partial r}\Big(r^2\frac{\partial}{\partial r}\Big)-\frac{\Lambda_2^2}{r^2}$, where 
$\Lambda_2^2$ can be obtained form Eq.(3.8) as $\Lambda_2^2=-\Big[\frac{1}{sin\theta}\frac{\partial}{\partial\theta}(sin\theta\frac{\partial}{\partial \theta})+\frac{1}{sin^2\theta}\frac{\partial^2}{\partial\varphi^2}\Big]$. So this concludes the verification that
\begin{eqnarray*}
\nabla^2=\frac{1}{r^2}\frac{\partial}{\partial r}\Big(r^2\frac{\partial}{\partial r}\Big)+\frac{1}{sin\theta}\frac{\partial}{\partial\theta}(sin\theta\frac{\partial}{\partial \theta})+\frac{1}{sin^2\theta}\frac{\partial^2}{\partial\varphi^2}\,.
\end{eqnarray*}   
Now we are in a position to write the $N$-dimensional fractional time independent Schr\"{o}dinger equation for a diatomic molecule of reduced mass (centre of mass coordinate system) $M=\frac{m_1m_2}{m_1+m_2}$ where $m_1$ and $m_2$ are the masses of constituent particles forming the molecule. If we choose the natural unit $\hbar=c=1$ then the form of the equation in large-$N$ expansion [33] will be
\begin{eqnarray}
\Big[\nabla_{N}^{2\alpha}+2M(\mathcal{E}_{\alpha}-V(r^{\alpha}))\Big]\psi(r^{\alpha},\Omega^{\alpha}_N)=0\,,
\end{eqnarray} 
where $\mathcal{E}_{\alpha}$ and $V(r^{\alpha})$ are the fractional energy and potential respectively. The term $\Omega^{\alpha}_N$ within the argument of $\psi$ denotes angular variables $\theta_1^{\alpha}, \theta_2^{\alpha}, \theta_3^{\alpha} \cdots \theta_{N-2}^{\alpha}, \phi^{\alpha}$. Taking the solution by means of separation variable technique 
$\psi(r^{\alpha},\Omega^{\alpha}_N)=R(r^{\alpha})Y(\Omega^{\alpha}_N)$ and adopting the eigenvalue equation for $Y(\Omega^{\alpha}_N)$ as (see Appendix-1 for more)
\begin{eqnarray}
\Lambda^{2\alpha}_{N-1}Y(\Omega^{\alpha}_N)=\ell(\ell+N-2)|_{N>1}Y(\Omega^{\alpha}_N)\,,
\end{eqnarray}
where $\ell$ is orbital angular momentum quantum number, we have the fractional order hyperradial or in short `radial' equation 
\begin{eqnarray}
\Bigg[\frac{d^{2\alpha}}{dr^{2\alpha}}+\frac{\Gamma(1+\alpha(N-1))}{\Gamma(1+\alpha(N-2))}\frac{1}{r^{\alpha}}\frac{d^{\alpha}}{dr^{\alpha}}-\frac{\ell(\ell+N-2)(\alpha!)^2}{r^{2\alpha}}\nonumber\\+2M(\alpha!)^2(\mathcal{E}_{\alpha}-V(r^{\alpha}))\Bigg]R(r^{\alpha})=0\,.
\end{eqnarray}
Here in deriving Eq.(3.11) from Eq.(3.9), we have expanded Eq.(3.7) by means of Jumarie derivative rules. It is worth to mention that, $\ell$ can take quantized values $0,1,2,3\cdots$ only.  
\section{B\lowercase{ound state spectrum of fractional} M\lowercase{ie-type potential}}
Inserting the potential (1.1) into Eq.(3.11) and using the following abbreviations:
\begin{subequations}
\begin{align}
\nu_{\alpha}(\nu_{\alpha}+1)&=\ell(\ell+N-2)(\alpha!)^2+2M(\alpha!)^2 A\,,\\
-\epsilon_{\alpha}^2&=2M(\alpha!)^2(\mathcal{E}_{\alpha}-C)\,,\\
-\beta_{\alpha}&=2M(\alpha!)^2 B\,,
\end{align}
\end{subequations}
we have
\begin{eqnarray}
\,_0^{J}D_{r}^{2\alpha}[R(r^{\alpha})]+\frac{\Gamma(1+\alpha(N-1))}{\Gamma(1+\alpha(N-2))}\frac{1}{r^{\alpha}}\,_0^{J}D_{r}^{\alpha}[R(r^{\alpha})]+\Big[\frac{\beta_{\alpha}}{r^{\alpha}}-\frac{\nu_{\alpha}(\nu_{\alpha}+1)}{r^{2\alpha}}-\epsilon_{\alpha}^2\Big]R(r^{\alpha})=0\,.
\end{eqnarray}
We choose the bound state eigenfunctions $\psi(r^{\alpha},\Omega^{\alpha}_N)$ that are vanishing for $r\rightarrow 0$ and $r\rightarrow \infty$. Since this type of initial conditions are associated with $R(r^{\alpha})$, we take the 
predetermined solutions of Eq.(4.2) as 
\begin{eqnarray}
R(r^{\alpha})=(r^{\alpha})^{-k}f(r^{\alpha})|_{k>0}\,.
\end{eqnarray}  
Here the term $(r^{\alpha})^{-k}$ ensures the fact that $R(r\rightarrow \infty)=0$. The unknown function $f(r^{\alpha})$ is expected to behave like $f(r\rightarrow 0)=0$. After deriving $\,_0^{J}D_{r}^{2\alpha}[f(r^{\alpha})]$, $\,_0^{J}D_{r}^{\alpha}[f(r^{\alpha})]$ and performing little calculation on Eq.(4.2) we have
\begin{eqnarray}
\,_0^{J}D_{r}^{2\alpha}f(r^{\alpha})+\frac{Q_{1}(\alpha,k,N)}{r^{\alpha}}\,_0^{J}D_{r}^{\alpha}f(r^{\alpha})+\Big[\frac{Q_{2}(\alpha,k,N,\nu_{\alpha})}{r^{2\alpha}}+\frac{\beta_{\alpha}}{r^{\alpha}}-\epsilon_{\alpha}^2\Big]f(r^{\alpha})=0\,,
\end{eqnarray}
where
\begin{subequations}
\begin{align}
 Q_{1}(\alpha,k,N)&=\frac{2\Gamma(1-\alpha k)}{\Gamma(1-\alpha k-\alpha)}+\frac{\Gamma(1+\alpha(N-1))}{\Gamma(1+\alpha(N-2))}\,,\\
 Q_{2}(\alpha,k,N,\nu_{\alpha})&=\frac{\Gamma(1-\alpha k)}{\Gamma(1-\alpha k-\alpha)}\Bigg[\frac{\Gamma(1-\alpha k-\alpha)}{\Gamma(1-\alpha k-2\alpha)}+\frac{\Gamma(1+\alpha(N-1))}{\Gamma(1+\alpha(N-2))}\Bigg]-\nu_{\alpha}(\nu_{\alpha}+1)\,.
\end{align}
\end{subequations}
Finding the solution of Eq.(4.4) is a difficult task due to the strong singular term $\frac{Q_{2}(\alpha,k,N,\nu_{\alpha})}{r^{2\alpha}}$. To ease out the situation we will study the Eq.(4.4) in transformed space (Laplace) with a parametric restriction 
\begin{eqnarray}
Q_{2}(\alpha,k,N,\nu_{\alpha})=0\,.
\end{eqnarray}
It is important to mention here that, above condition is not a mandatory or essential to apply the Laplace transform on Eq.(4.4). The imposed condition only helps to reduce the tenacious mathematical steps. Denoting the solution of Eq.(4.6) for $k(>0)$ as $k_{\alpha}^{*}$, we can rewrite Eq.(4.4) as
\begin{eqnarray}
r^{\alpha}\,_0^{J}D_{r}^{2\alpha}g(r)+Q_{1}(\alpha,k_{\alpha}^{*},N)\,_0^{J}D_{r}^{\alpha}g(r)+(\beta_{\alpha}-\epsilon_{\alpha}^2r^{\alpha})g(r)=0\,,
\end{eqnarray} 
where $f(r^{\alpha})$ is replaced with $g(r)$. Now defining
$\mathcal{L}\left\{g(r)\right\}=\zeta(s)$ and using the rules of Laplace transform, mentioned in subsection (2.2) with $g(0)=0$, it is easy to obtain the following fractional differential equation (see Appendix-2)
\begin{eqnarray}
\,_0^{J}D_{s}^{\alpha}\zeta(s)+\eta(s^{\alpha})\zeta(s)=0\,,
\end{eqnarray}   
where 
\begin{subequations}
\begin{align}
\eta{(s^{\alpha})}&=\frac{\lambda_1}{s^{\alpha}+\epsilon_{\alpha}}+\frac{\lambda_2}{s^{\alpha}-\epsilon_{\alpha}}\,,\\
\lambda_1&=\frac{1}{2}\Big[\gamma_{\alpha}+\frac{\beta_{\alpha}}{\tau\epsilon_{\alpha}}\Big]\,,\\
\lambda_2&=\frac{1}{2}\Big[\gamma_{\alpha}-\frac{\beta_{\alpha}}{\tau\epsilon_{\alpha}}\Big]\,,\\
\gamma_{\alpha}&=\frac{\Gamma(1+2\alpha)}{\Gamma(1+\alpha)}-\frac{Q_{1}(\alpha,k_{\alpha}^{*},N)}{\tau}\,.
\end{align}
\end{subequations}
Exact solution of Eq.(4.8) is very complicated and tedious in fractional domain. The good news is we can approximate the solution very near to the $\alpha\approx1.00$ (see Appendix-2). Adopting Jumarie sense integration with the definition of `$\alpha$-logarithmic' function [34]i.e, $\int\frac{d^\alpha t}{t}=Ln_\alpha(\frac{t}{C}), t=E_\alpha(Ln_\alpha t)$ where $C$ denotes a constant such that $(\frac{t}{C})>0$, we have the approximate solution in transformed space as
\begin{eqnarray}
\zeta(s)=C_{1}(s^{\alpha}+\epsilon_{\alpha})^{-\gamma_{\alpha}}\Big(\frac{s^{\alpha}-\epsilon_{\alpha}}{s^{\alpha}+\epsilon_{\alpha}}\Big)^{-\lambda_2}\,,
\end{eqnarray}
where $C_1$ is the integration constant. The second factor of Eq.(4.10) is a multivalued function when the power $-\lambda_2$ is a non integer. The quantum mechanical eigenfunction must be single valued in nature. So we must take
\begin{eqnarray}
-\lambda_2=\frac{1}{2}\Big[\frac{\beta_{\alpha}}{\tau\epsilon_{\alpha}}-\gamma_{\alpha}\Big]=n \,, n=0,1,2,3, \ldots
\end{eqnarray}   
The inverse transform of Eq.(4.10) will provide the solution of the problem in actual space. To that aim, we expand Eq.(4.10) with help of Eq.(4.11) as  
\begin{align}
\zeta(s)&=C_1(s^{\alpha}+\epsilon_{\alpha})^{-\gamma_{\alpha}}\Big[1-\frac{2\epsilon_{\alpha}}{s^{\alpha}+\epsilon_{\alpha}}\Big]^{n}\,, \nonumber \\ 
&=C_1\sum_{j=0}^{n}\frac{n!}{j!(n-j)!}(-1)^{j}(2\epsilon_{\alpha})^{j}(s^{\alpha}+\epsilon_{\alpha})^{-(\gamma_{\alpha}+j)}\,,\nonumber \\
&=C_1\sum_{j=0}^{n}\frac{n!}{j!(n-j)!}(-1)^{j}(2\epsilon_{\alpha})^{j}\frac{1}{(s^{\alpha}+\epsilon_{\alpha})^{m_j+1}}\,, \quad\text{where $m_j=(\gamma_{\alpha}+j-1)$}\,.
\end{align}
Using the formula given by Eq.(2.11d) for $\alpha=\beta$ we can find the inverse of Eq.(4.12) quite easily. 
\begin{align}
g(r)&=C_1\sum_{j=0}^{n}\frac{n!}{j!(n-j)!}(-1)^{j}(2\epsilon_{\alpha})^{j} \frac{1}{\Gamma(\gamma_{\alpha}+j)}r^{\alpha(\gamma_{\alpha}+j)-1}E_{\alpha}^{(m_j)}(-\epsilon_{\alpha}r^{\alpha})\,, \nonumber\\
&=\frac{C_1}{\Gamma(\gamma_{\alpha})}r^{\alpha\gamma_{\alpha}-1}\sum_{j=0}^{n}\frac{n!}{j!(n-j)!}(-1)^{j} \frac{\Gamma(\gamma_{\alpha})}{\Gamma(\gamma_{\alpha}+j)}(2\epsilon_{\alpha}r^{\alpha})^{j}E_{\alpha}^{(m_j)}(-\epsilon_{\alpha}r^{\alpha})\,,\nonumber\\
 &=\mathcal{N}_cr^{\alpha\gamma_{\alpha}-1}E_{\alpha}^{(m_j)}(-\epsilon_{\alpha}r^{\alpha})\,_{1}F_{1}(-n,\gamma_{\alpha}, 2\epsilon_{\alpha}r^{\alpha})\,,
\end{align} 
where $E_{\alpha}^{(m_j)}(-\epsilon_{\alpha}r^{\alpha})=\frac{d^{m_j}}{dr^{m_j}}E_{\alpha}(-\epsilon_{\alpha}r^{\alpha})=\sum_{p=0}^{\infty}\frac{(p+m_j)!}{p!}\frac{(-\epsilon_{\alpha}r^{\alpha})^{p}}{\Gamma(\alpha p+\alpha m_j+\alpha)}$[19] and $\,_{1}F_{1}$ is fractionally defined confluent hypergeometric function i.e 
\begin{eqnarray*}
\,_{1}F_{1}(-n,\gamma_{\alpha}, 2\epsilon_{\alpha}r^{\alpha})=\sum_{j=0}^{n}\frac{n!}{j!(n-j)!}(-1)^{j} \frac{\Gamma(\gamma_{\alpha})}{\Gamma(\gamma_{\alpha}+j)}(2\epsilon_{\alpha}r^{\alpha})^{j}\,.
\end{eqnarray*}  
Hence the complete radial eigenfunctions are
\begin{eqnarray}
R_{n\alpha N \ell}(r^{\alpha})=r^{-\alpha k_{\alpha}^{*}}g(r)=\mathcal{N}_c r^{\alpha(\gamma_{\alpha}-k_{\alpha}^{*})-1}E_{\alpha}^{(m_j)}(-\epsilon_{\alpha}r^{\alpha})\,_{1}F_{1}(-n,\gamma_{\alpha}, 2\epsilon_{\alpha}r^{\alpha})\,,
\end{eqnarray}  
where $\mathcal{N}_c=\frac{C_1}{\Gamma(\gamma_{\alpha})}$ acts like a normalization constant. The energy eigenvalue equation of the potential model comes out from Eq.(4.11) as
\begin{eqnarray}
\mathcal{E}_{n\alpha N \ell}=C-\frac{M(\alpha!)^2}{2\tau^2}\Bigg[\frac{B}{n+\frac{1}{2}\frac{\Gamma(1+2\alpha)}{\Gamma(1+\alpha)}-\frac{Q_1(\alpha,k^{*}_\alpha, N)}{2\tau}}\Bigg]^2\,.
\end{eqnarray} 
\section{D\lowercase{iscussion}}
Originally, Schr\"{o}dinger equation is a second order differential equation with non-constant coefficients under the potential model $V(r)$. In this study we have generalized the potential model as $V(r^{\alpha})$ and eventually studied fractional Schr\"{o}dinger equation in multidimensional space. As an example generalized Mie-type potential has been taken as $V(r^{\alpha})$ and solved approximately for $\alpha\approx 1.00$. The solution is vary closed to the exact one, which means we can express the grand solution as   
\begin{eqnarray*}
\Psi(r^{\alpha},\Omega^{\alpha}_N)=\sum_n \mathcal{C}_n\psi_n(r^{\alpha},\Omega^{\alpha}_N)\,,
\end{eqnarray*} 
where $\sum_n$ helps to express the overall solution in terms of all possible solutions i.e a linear combinational form via the constants $C_n(n=1,2,3...)$. It has been noticed that for a particular value of $\alpha(0<\alpha<1)$ the solution depends on the dimensionality as well as on the potential variables $A,B,C$. The potential parameter $B$ is not a free parameter as $A$ and $C$. For example, under any circumstances the parameter $B$ can not be taken as zero otherwise it will force $\beta_{\alpha}$ to have a zero value and spoil the quantization condition given by Eq.(4.11). Here we have few special cases   

\subsection{\textit{Fractional Coulomb potential}}
For this case $A=C=0$ and $B\neq 0$ make the potential as $V(r^{\alpha})=\frac{B}{r^{\alpha}}$. Immediately we have the energy eigenvalue equation
\begin{eqnarray}
\mathcal{E}_{n\alpha N \ell}=-\frac{M(\alpha!)^2}{2\tau^2}\Bigg[\frac{B}{n+\frac{1}{2}\frac{\Gamma(1+2\alpha)}{\Gamma(1+\alpha)}-\frac{Q_1(\alpha,k^{*}_\alpha, N)}{2\tau}}\Bigg]^2\,, 
\end{eqnarray}
with radial eigenfunction 
\begin{eqnarray}
R_{n\alpha N \ell}(r^{\alpha})=\mathcal{N}_c r^{\alpha(\gamma_{\alpha}-k_{\alpha}^{*})-1}E_{\alpha}^{(m_j)}(-\epsilon_{\alpha}r^{\alpha})\,_{1}F_{1}(-n,\gamma_{\alpha}, 2\epsilon_{\alpha}r^{\alpha})\,.
\end{eqnarray}
In this case the values of $k_{\alpha}^{*}$ emerge from the Eq.(4.6) where $\nu_{\alpha}(\nu_{\alpha}+1)=\ell(\ell+N-2)(\alpha!)^2$.
Now when $\alpha=1$ (as a result $\tau=1$), the Eq.(4.6) provides $k_{1}^{*}=\ell+N-2$ and hence $Q_{1}(1,k_{1}^{*},N)=-2\ell-N+3$. The energy eigenvalue becomes
\begin{eqnarray}
\mathcal{E}_{n1N\ell}=-\frac{M}{2}\Bigg[\frac{B}{n+\ell+\frac{N-1}{2}}\Bigg]^2\,.
\end{eqnarray}
The radial eigenfunctions come out as
\begin{eqnarray}
R_{n1N\ell}(r)=\mathcal{N^{'}} r^{\ell} e^{-\epsilon r}\,_{1}F_{1}(-n,2\ell+N-1,2\epsilon r)\,,
\end{eqnarray}
where $\mathcal{N^{'}}$ is new normalization constant viz $\mathcal{N^{'}}=\mathcal{N}_c(-\epsilon)^{m_j}$  
The results are similar with the previous works [19,35]. 
\subsection{\textit{Mie-type potential in $N$ dimension when $\alpha=1$}}
Under this condition it is easy to rewrite Eq.(4.6) as $k(k+1)-k(N-1)-\nu(\nu+1)=0$. For simplicity we write the positive solution of $k$ as $k_1^{*}=k_{\ell N}$. This enable us to find $Q_{1}(1,k_{1}^{*},N)=-2k_{\ell N}+N-1$. Hence the energy eigenvalues for this situation emerge as
\begin{eqnarray}
\mathcal{E}_{n1N\ell}=C-\frac{M}{2}\Bigg[\frac{B}{n+k_{\ell N}+\frac{3-N}{2}}\Bigg]^2\,,
\end{eqnarray}
with the radial eigenfunctions 
\begin{eqnarray}
R_{n1N\ell}(r)=\mathcal{N^{'}} r^{k_{\ell N}+2-N} e^{-\epsilon r}\,_{1}F_{1}(-n,2k_{\ell N}-N+3,2\epsilon r)\,.
\end{eqnarray}
$\mathcal{N^{'}}$ acts the normalization constant as previous. These all outcomes in this subsection are well matched with the results [36].\\
Apart from the theoretical aspects, we have examined the present model numerically also. For a typical diatomic molecule ($M=0.31GeV, D_0=2\times 10^{-9} GeV, r_0=10^5 GeV^{-1}$) energy eigenvalues have been derived in table-1. This clearly shows bound state energy ($E<0$) is possible for $\alpha(0<\alpha<1)$ to some extent. In case of dimension $N=3$, physically significant energy eigenvalues come out in the range $0.85<\alpha<1$ for $n=1,2$ states with $\ell=1$. Corresponding eigenfunctions have been shown in FIG.1 and FIG.2. The eigenfunctions are continuous and well behaved, but for lower 
$\alpha$, such as near $\alpha=0.85$, the graph looses its periodicity significantly and tries to blow up in the selected range. As the dimension $N$ increases, the quantum mechanical well behaved eigenfunctions are only possible from 
$\alpha=0.95$ to $\alpha=1$. These are displayed in FIG.3,FIG.4,FIG.5,FIG.6 respectively. In these cases, going to the further lower $\alpha$ near $0.90$ the eigenfunction losses its well behaved property for $N=4,5$. This is due to the value of the $k_{\alpha}^*$ originated from the condition $Q_{2}(\alpha,k,N,\nu_{\alpha})=0$. It is important to mention here that, the function $Q_2$ is very crucial to stabilize the entire model of the present study. The solution of 
$Q_{2}(\alpha,k,N,\nu_{\alpha})=0$, i.e $k_{\alpha}^*$, has several possibilities but only bound states are possible for those $k_{\alpha}^*$ for which the difference $(\gamma_{\alpha}-k_{\alpha}^*)$ is not so large to make the factor 
$r^{\alpha(\gamma_{\alpha}-k_{\alpha}^{*})-1}$, in Eq.(4.14), a sharp increasing one. 
\begin{table}[http]
\begin{center}
{\bf Fractional Kartzer-Fues potential}\\
Different potential parameters for a typical diatomic molecule and $\tau$ against $\alpha$ 
\caption{$D_0=2\times 10^{-9}GeV$, $M=0.31 GeV$, $r_0=10^5 GeV^{-1}$, $C=0$, $\delta=-0.5$, $\ell=1$}
\renewcommand{\arraystretch}{0.5}
\begin{tabular}{|>{\centering\arraybackslash}m{1in}|>{\centering\arraybackslash}m{1in}|>{\centering\arraybackslash}m{1in}|>{\centering\arraybackslash}m{1in}|}\hline
$\alpha$ & $ A=D_0r_0^{2\alpha}$ & $ B=-2D_0r_0^{\alpha}$& $\tau$\\ \hline
$0.70$ & $0.0200$ & $-1.2649\times 10^{-5}$ & $1.7013$ \\ \hline
$0.75$ & $0.0632$ & $-2.2494\times 10^{-5}$ & $1.4142$ \\ \hline
$0.80$ & $0.2000$ & $-4.0\times 10^{-5}$ & $1.2361$ \\ \hline
$0.85$ & $0.6325$ & $-7.1131\times 10^{-5}$ & $1.2223$ \\ \hline
$0.90$ & $2.000$ & $-1.2649\times 10^{-4}$ & $1.0515$ \\ \hline
$0.95$ & $6.3246$ & $-2.2494\times 10^{-4}$ & $1.0125$ \\ \hline
$1.0$ & $20.0$ & $-4.0\times 10^{-4}$ & $1.0$ \\ \hline
\end{tabular}
\end{center}
\end{table} 
\begin{table}[htbb]
\begin{center}
Energy spectrum of the Molecule (eV unit) 
\renewcommand{\arraystretch}{0.4}
\centerline{
\begin{tabular}{|>{\centering\arraybackslash}m{1in}|>{\centering\arraybackslash}m{1in}|>{\centering\arraybackslash}m{1in}|>{\centering\arraybackslash}m{1in}|>{\centering\arraybackslash}m{1in}|>{\centering\arraybackslash}m{1in}|>{\centering\arraybackslash}m{1in}|}\hline
$N$ & $\alpha$ &$k_{\alpha}^*$ & $Q_1$& $\gamma_\alpha$ &$\mathcal{E}(n=1)$& $\mathcal{E}(n=2)$ \\ \hline
&$0.70$ & $1.401540$ & $-24.0199$ &$15.4856$ & $-8.9562\times 10^{-5}$ &$-7.212\times 10^{-5}$ \\ \cline{2-7}
&$0.75$ & $1.277020$ & $-9.2051$ &$7.9555$ & $-0.0013$ &$-8.9706\times 10^{-4}$ \\ \cline{2-7}
&$0.80$ & $2.280294$ & $-3.9385$ &$4.7212$ & $-0.0121$ &$-0.0072$ \\ \cline{2-7}
$3$&$0.85$ & $2.118725$ & $-3.2152$ &$4.2640$ & $-0.0463$ &$-0.0266$ \\ \cline{2-7}
&$0.90$ & $2.030419$ & $-3.1805$ &$4.7679$ & $-0.1753$ &$-0.1045$ \\ \cline{2-7}
&$0.95$ & $5.452018$ & $-5.1659$ &$6.9670$ & $-0.3536$ &$-0.2364$ \\ \cline{2-7}
&$1.0$ & $4.327531$ & $-6.6551$ &$8.6551$ & $-0.8456$ &$-0.5994$ \\ \cline{1-7}
&$0.70$ & $8.293139$ & $-9.9419$ &$7.2108$ & $-3.2277\times 10^{-4}$ &$-2.1788\times 10^{-4}$ \\ \cline{2-7}
&$0.75$ & $1.324563$ & $-61.9221$ &$45.2324$ & $-5.7474\times 10^{-5}$ &$-5.29\times 10^{-5}$ \\ \cline{2-7}
&$0.80$ & $2.382292$ & $-6.7296$ & $6.9792$ & $-0.0068$ & $-0.0045$ \\ \cline{2-7}
$4$&$0.85$ & $6.463783$ & $-5.1534$ &$5.8497$ & $-0.0295$ &$-0.0187$ \\ \cline{2-7}
&$0.90$ & $2.128283$ & $-4.8426$ &$6.3486$ & $-0.1152$ &$-0.0750$ \\ \cline{2-7}
&$0.95$ & $5.529740$ & $-5.3053$ &$7.1047$ & $-0.3430$ &$-0.2306$ \\ \cline{2-7}
&$1.0$ & $5.0496913$ & $-7.0994$ &$9.0994$ & $-0.7792$ &$-0.5595$ \\ \cline{1-7}
&$0.70$ & $8.352522$ & $-12.3138$ &$8.6049$ & $-2.4348\times 10^{-4}$ &$-1.7235\times 10^{-4}$ \\ \cline{2-7}
&$0.75$ & $7.687194$ & $-8.5297$ &$7.4779$ & $-0.0014$ &$-9.7326\times 10^{-4}$ \\ \cline{2-7}
&$0.80$ & $7.095835$ & $-6.6828$ &$6.9413$ & $-0.0068$ &$-0.0046$ \\ \cline{2-7}
$5$&$0.85$ & $4.425374$ & $-5.4828$ &$6.1192$ & $-0.0276$ &$-0.0177$ \\ \cline{2-7}
&$0.90$ & $7.1076947$ & $-5.7343$ &$7.1966$ & $-0.0950$ &$-0.0641$ \\ \cline{2-7}
&$0.95$ & $5.6810119$ & $-5.6611$ &$7.4561$ & $-0.3180$ &$-0.2167$ \\ \cline{2-7}
&$1.0$ & $5.818564$ & $-7.6371$ &$9.6371$ & $-0.7089$ &$-0.5162$ \\ \cline{1-7}
\end{tabular}
}
\end{center}
\end{table} 
\begin{figure}[http]
	\centering
		\includegraphics[width=0.75\textwidth]{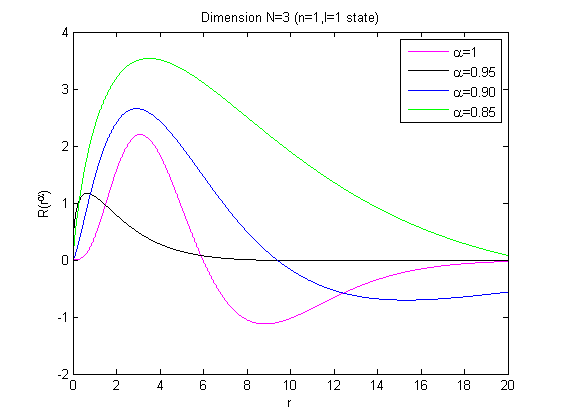}
	\caption{n=1 state eigenfunctions in N=3 dimension for fractional Kartzer-Fues potential @ $\alpha=1.0,0.95,0.90,0.85$ }
	\label{fig:Fig1}
\end{figure}
\begin{figure}[http]
	\centering
		\includegraphics[width=0.75\textwidth]{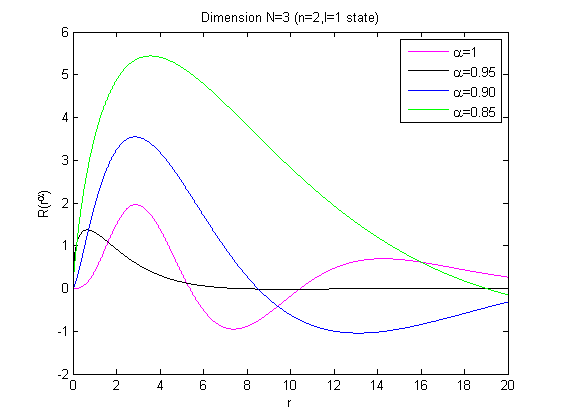}
	\caption{n=2 state eigenfunctions in N=3 dimension for fractional Kartzer-Fues potential @ $\alpha=1.0,0.95,0.90,0.85$}
	\label{fig:Fig2}
\end{figure}
\begin{figure}[http]
	\centering
		\includegraphics[width=0.75\textwidth]{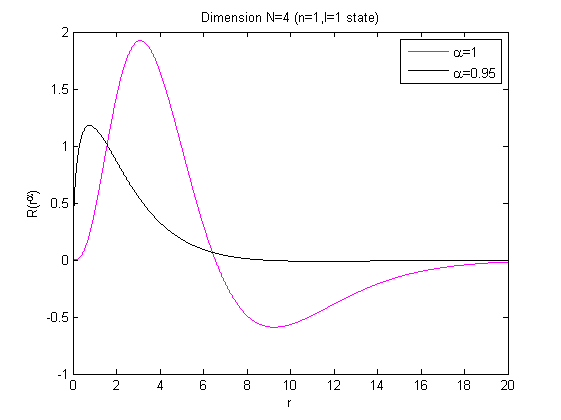}
	\caption{n=1 state eigenfunctions in N=4 dimension for fractional Kartzer-Fues potential @ $\alpha=1.0,0.95$}
	\label{fig:Fig3}
\end{figure}
\begin{figure}[http]
	\centering
		\includegraphics[width=0.75\textwidth]{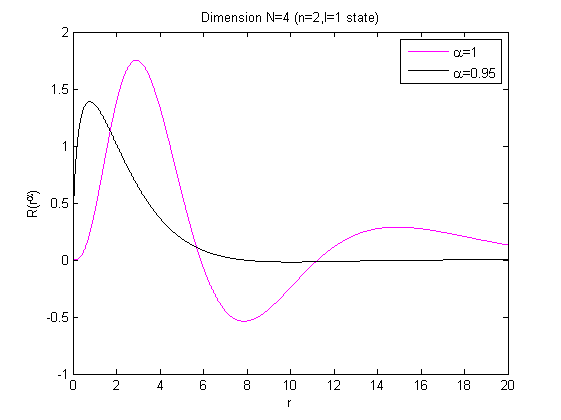}
	\caption{n=2 state eigenfunctions in N=4 dimension for fractional Kartzer-Fues potential @ $\alpha=1.0,0.95$}
	\label{fig:Fig4}
\end{figure}
\begin{figure}[http]
	\centering
		\includegraphics[width=0.75\textwidth]{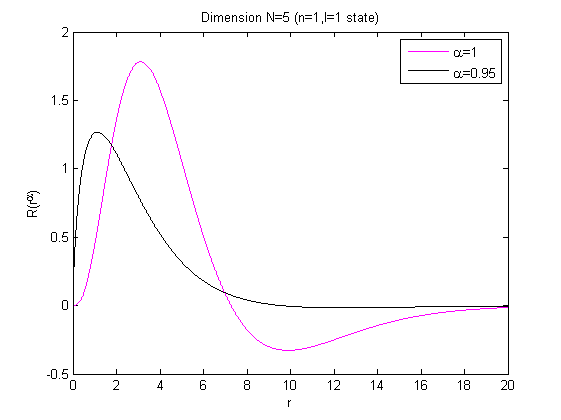}
	\caption{n=1 state eigenfunctions in N=5 dimension for fractional Kartzer-Fues potential @ $\alpha=1.0,0.95$}
	\label{fig:Fig5}
\end{figure}
\begin{figure}[http]
	\centering
		\includegraphics[width=0.75\textwidth]{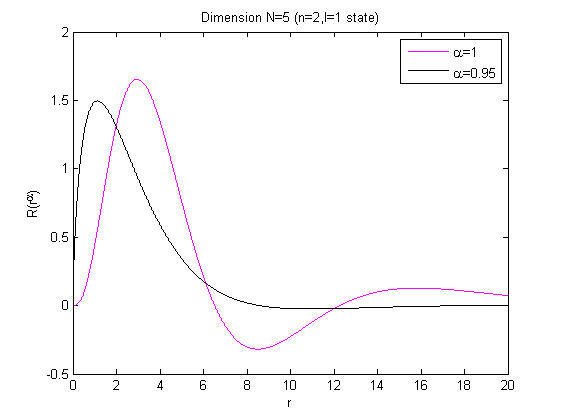}
	\caption{n=2 state eigenfunctions in N=5 dimension for fractional Kartzer-Fues potential @ $\alpha=1.0,0.95$}
	\label{fig:Fig6}
\end{figure}
\begin{figure}[http]
	\centering
		\includegraphics[width=0.75\textwidth]{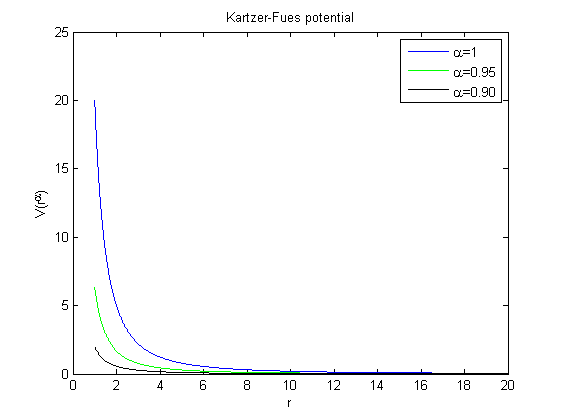}
	\caption{Kartzer-Fues potential variation with $r$ }
	\label{fig:Fig7}
\end{figure}
\section{C\lowercase{onclusion}}
In this paper, we have studied approximate bound state solutions of $N$-dimensional fractional Schr\"{o}dinger equation for generalised Mie-type potential, namely $V(r^{\alpha})=\frac{A}{r^{2\alpha}}+\frac{B}{r^{\alpha}}+C$, where $\alpha(0<\alpha<1)$ acts like a fractional parameter. We have framed the entire study by Jumarie type derivative rules. Applying the separation variable method, $N$-dimensional fractional Schr\"{o}dinger equation has been separated into hyperradial(or radial) and hyperangular equations which contain the fractional order derivatives. The radial fractional order differential equation, coupled with the eigenvalues of hyperangular part, is found complicated due to the strong singular term $\frac{1}{r^{2\alpha}}$. To solve this, after introducing a parametric restriction to remove the strong singular term, we have constructed the replica of radial equation in transformed space (Laplace space) which is comparatively easy to tackle, as it contains the lower order fractional derivative of transformed variable. Inverse transform of the transformed equation provides us the radial eigenfunction in actual space with one parameter Mittag-Leffler function and fractional confluent hypergeometric function. Energy eigenvalue has been obtained via a quantization condition which also prevent the eigenfunction to turned into a multivalued one. The results are verified for (i) Coulomb potential in $N$ and $3$ dimensions (ii) Mie-type potential in $3$ dimensions. We have also furnished numerical results as well as few eigenfunctions for typical diatomic molecular system and studied their variation with $\alpha$. This present study encourages to use fractional Schr\"{o}dinger equation for different potential models and disclose the hidden physics behind it for $0<\alpha<1$. Before ending the conclusion we have drawn FIG.7, the Kartzer-Fues potentials for different $\alpha$. There is an interesting physics lying beneath of the FIG.7 and associated eigenfunctions. As $\alpha$ is lowered the strength of the potential becomes smaller i.e tending to zero. In other word we can interpret that the under examined particle is somehow becoming a free particle type. Under such circumstance, the eigenfunctions specially for $(N>3)$ and lower $\alpha$ tend to approach to the eigenfunction of a infinite spherical well type problem or the quantum mechanical box problem of a free particle. Though the periodicity losses, the spatial spreading of the eigenfunctions become narrower eventually that signifies that the probability of finding the particle in the specified range becomes higher. Now since the uncertainty in position of the particle decreases, according to uncertainty principle the uncertainty in momentum will increase considerably. This will also make the energy more uncertain. This is what we have achieved in TABLE 1. It is clear in table that as we go to the higher dimension with lower $\alpha$ the energy eigenvalues become less prominent. More work in the aspect of uncertainty principle via fractional Schr\"{o}dinger equation need to be carried out.

\newpage
\section*{R\lowercase{eferences}}

\newpage
\small
\renewcommand{\theequation}{A1-\arabic{equation}}
\setcounter{equation}{0}  
\section*{Appendix-1}
{\bf A1-1: N dimensional space in ordinary sense}\\
The relations between the Cartesian coordinates $x_i$ and the hyperspherical coordinates $r$ and $\theta_b$ in $N$ dimensional space are defined 
\begin{align}
x_1 &= rcos\theta_1sin\theta_2 \cdots sin\theta_{N-1}\nonumber\,,\\
x_2 &= r sin\theta_1sin\theta_2 \cdots sin\theta_{N-1}\nonumber\,,\\
x_b &= r cos\theta_{b-1}sin\theta_b \cdots sin\theta_{N-1}\nonumber\,,\\
x_N &= r cos\theta_{N-1}\nonumber\,,
\end{align}
where $b\in[3,N-1]$. These generate $r^2=\sum_{i=1}^N x_i^2$. The volume element of the configuration space is calculated  as $\prod_{a=1}^Ndx_a=r^{N-1}dr d\Omega$ where $\Omega=\prod_{a=1}^{N-1}(sin\theta_a)^{a-1}d\theta_a$ with $r\in(0,\infty), \theta_1\in(-\pi.\pi)$ and $\theta_c\in(0,\pi), c\in(2,N-1)$. \\
{\bf A1-2: The hyperspherical harmonics $Y(\Omega^{\alpha}_N)$}\\
Ordinary spherical harmonic $Y(\theta,\varphi)$, in case of Schr\"{o}dinger Hydrogen atom problem, is eigenfunction of angular momentum operator $\Lambda^2$. The eigenvalue is characterized by the angular momentum quantum number $\ell$ and dimension $N$. We know in two dimension the eigenvalue of angular momentum operator is $\ell^2\equiv \ell(\ell+2-2)$, for three dimension it is $\ell(\ell+1)\equiv \ell(\ell+3-2)\cdots$  So in $N$ dimension the eigenvalue is taken as $\ell(\ell+N-2)$. As $\ell$ is a quantum number it is restricted to take integer values like $0,1,2,3,\cdots$.
In fractional Schr\"{o}dinger equation, for radial symmetry fractional potential problem in $N$ dimension, the spherical harmonic $Y(\Omega^{\alpha}_N)$ is named as hyperspherical harmonic via fractional generalised coordinates. In this study we proposed that the eigenvalue equation for generalised angular momentum operator is of the form
\begin{eqnarray}
\Lambda^{2\alpha}Y(\Omega^{\alpha}_N)=\ell(\ell+N-2)Y(\Omega^{\alpha}_N)\,,
\end{eqnarray}
where eigenvalue $\ell(\ell+N-2)$ is independent of fractional parameter $\alpha$.
   
\section*{Appendix-2}
\renewcommand{\theequation}{A2-\arabic{equation}}
\setcounter{equation}{0}  
\enumerate
\item {\bf Proof of Eq.(2.9b)}\\
let us take $\phi(s)=\mathcal{L}\left\{f(x)\right\}=s^\delta$ with $-1<\delta<0$. Then
\begin{align}
\frac{d^\alpha}{ds^\alpha}\mathcal{L}\left\{f(x)\right\}&=\frac{d^\alpha[s^\delta]}{ds^\alpha}\nonumber\\
&=\frac{\Gamma(\delta+1)}{\Gamma(\delta-\alpha+1)}s^{\delta-\alpha}\nonumber\\
&=\frac{\Gamma(\delta+1)}{\Gamma(\delta-\alpha+1)}\mathcal{L}\left\{\frac{x^{\alpha-\delta-1}}{\Gamma(\alpha-\delta)}\right\}\nonumber\\
&=\frac{\Gamma(\delta+1)\Gamma(-\delta)}{\Gamma(\delta-\alpha+1)\Gamma(\alpha-\delta)}\mathcal{L}\left\{x^\alpha f(x)\right\}\nonumber\\
&=\frac{cosec(-\delta\pi)}{cosec((\alpha-\delta)\pi)}\mathcal{L}\left\{x^\alpha f(x)\right\}\,,\nonumber
\end{align}
where we have used $\Gamma(-y)\Gamma(y+1)=-\pi cosec(\pi y)$ and $\mathcal{L}\left\{x^m\right\}=\frac{(m+1)}{s^{m+1}}$ in the above steps. So we can have
\begin{eqnarray}
\mathcal{L}\left\{x^\alpha f(x)\right\}=-\tau\frac{d^\alpha}{ds^\alpha}\mathcal{L}\left\{f(x)\right\}=-\tau\frac{d^\alpha}{ds^\alpha}\phi(s)\,,
\end{eqnarray}
where $\tau=-\frac{cosec((\alpha-\delta)\pi)}{cosec(-\delta\pi)}$ and when $\alpha=1$ it makes $\tau=1$. \\
The above relation is valid for a function $f(x)$ such that $f(x)=\frac{x^{-\delta-1}}{\Gamma(-\delta)}; \,\,\ \delta\neq 0$ for which the Laplace transform is $\phi(s)=s^\delta$. The above example suggest that it is futile to seek generalization of classical formula as in classical calculus \& Laplace transform i.e $\mathcal{L}\left\{x^{n}f(x)\right\}=(-1)^{n}\phi^{(n)}(s)$. However, for $n$ as non integer we use the expression $\mathcal{L}\left\{x^\alpha f(x)\right\}=-\tau\frac{d^\alpha}{ds^\alpha}\phi(s)$. A simple criterion for a function $f(x)\sim x^p$ is fractionally differintegrable when $p>-1$, comes form a fact that $D^\alpha x^p=\frac{\Gamma(p+1)}{\Gamma(p+1-\alpha)}x^{p-\alpha}$ to be finite one needs 
$\frac{\Gamma(p+1)}{\Gamma(p+1-\alpha)}$ should exist; that is when $p+1>0$ and $p+1-\alpha>0$, i.e we are excluding $p+1=0$ and $p+1-\alpha=0$, where the Gamma function is blowing out, With these two we say $p>-1$ as condition for a function $f(x)\sim x^p$ as fractionally differintegrable. For our case $\mathcal{L}\left\{f(x)\right\}=s^\delta$ is fractionally differintegrable if $\delta>-1$ and our
 $f(x)=\frac{x^{-\delta-1}}{\Gamma(-\delta)}=\mathcal{L}^{-1}\left\{s^\delta\right\}$, where $\delta\neq 0$,  thus we take $-1<\delta<0$ to have values of $\tau$.
\item{\bf Derivation of Eq.(4.10)}\\
Let us write the Eq.(4.7) in a general way and find the solution with given initial condition. So the problem is find the solution of
\begin{eqnarray}
x^{\alpha}\frac{d^{2\alpha}y(x)}{dx^{2\alpha}}+c_1\frac{d^{\alpha}y(x)}{dx^{\alpha}}+(c_2-c_3^2x^{\alpha})y(x)=0\,,\,\,c_i(i=1,2,3)\,\mbox{are constants}
\end{eqnarray} 
with the initial condition $y(0)=0$.\\
Let the Laplace transform of $y(x)$ is taken as $z(t)$ i.e $z(t)=\mathcal{L}\left\{y(x)\right\}$. Following section-2 we have these
\begin{subequations}
\begin{align}
\mathcal{L}\left\{\frac{d^{\alpha}y(x)}{dx^{\alpha}}\right\}&=t^\alpha z(t)\,,\\
\mathcal{L}\left\{x^{\alpha}y(x)\right\}&=-\tau\frac{d^\alpha z(t)}{dt^\alpha}\,,\\
\mathcal{L}\left\{x^\alpha\frac{d^{2\alpha}y(x)}{dx^{2\alpha}}\right\}&=-\tau z(t)\frac{\Gamma(1+2\alpha)}{\Gamma(1+\alpha)}t^\alpha-\tau t^{2\alpha}\frac{d^{\alpha}z(t)}{dt^{\alpha}}\,.
\end{align}
\end{subequations} 
Now taking the Laplace transform on Eq.(A2-2) and using above three results we have
\begin{eqnarray}
-\tau z(t)\frac{\Gamma(1+2\alpha)}{\Gamma(1+\alpha)}t^\alpha-\tau t^{2\alpha}\frac{d^{\alpha}z(t)}{dt^{\alpha}}+c_1t^\alpha z(t)+c_2z(t)+c_3^2 \tau\frac{d^{\alpha}z(t)}{dt^{\alpha}}=0\,.
\end{eqnarray}
Rearrangement of the transformed Equation (A2-4) yields
\begin{align}
(t^{2\alpha}-c_3^2)\frac{d^{\alpha}z(t)}{dt^{\alpha}}+(bt^{\alpha}-c)z(t)&=0\,,\nonumber\\
\frac{d^{\alpha}z(t)}{dt^{\alpha}}+\eta(t^\alpha)z(t)&=0\,.
\end{align}
where $b=\frac{\Gamma(1+2\alpha)}{\Gamma(1+\alpha)}-\frac{c_1}{\tau}$, $c=\frac{c_2}{\tau}$ and $\eta(t^\alpha)=\frac{bt^{\alpha}-c}{t^{2\alpha}-c_3^2}$. Now writing $\eta(t^\alpha)=\frac{\lambda_1}{t^\alpha+c_3}+\frac{\lambda_2}{t^\alpha-c_3}$ with $\lambda_1=\frac{b}{2}+\frac{c}{2c_3}$ and  $\lambda_2=\frac{b}{2}-\frac{c}{2c_3}$ Eq.(A2-5) can be expressed as 
\begin{eqnarray*}
\frac{d^\alpha z(t)}{z(t)}=-\Big(\frac{\lambda_1}{t^\alpha+c_3}+\frac{\lambda_2}{t^\alpha-c_3}\Big)dt^\alpha\,.
\end{eqnarray*}
Proceeding to the integration of the last step we can write
\begin{eqnarray}
\int\frac{d^\alpha z(t)}{z(t)}=-\Big(\int_0^t\frac{\lambda_1d\xi^\alpha}{\xi^\alpha+c_3}+\int_0^t\frac{\lambda_2d\xi^\alpha}{\xi^\alpha-c_3}\Big)\,.
\end{eqnarray}
Using the definition of fractional logarithmic function in Jumarie sense\\
$\int\frac{d^\alpha t}{t}=Ln_\alpha(\frac{t}{C}), t=E_\alpha(Ln_\alpha t)$ where $C$ denotes a constant such that
$(\frac{t}{C})>0$ and $Ln_\alpha t$ denotes the inverse function of the Mittag-Leffler function. The integral form of $Ln_\alpha t$ is given by $Ln_\alpha t=\frac{1}{(1-\alpha)!}\int_0^t(\frac{d\xi}{\xi})^\alpha$. So we have
\begin{eqnarray*}
\int\frac{d^\alpha z(t)}{z(t)}=Ln_\alpha\Big(\frac{z(t)}{C}\Big)\,.
\end{eqnarray*}
\begin{align}
\int\frac{dt^\alpha}{t^\alpha+c_3}&=\int_0^t\frac{d\xi^\alpha}{\xi^\alpha+c_3}\nonumber\\
&=\int_{\sqrt[\alpha]{c_3}}^{{\sqrt[\alpha]{t^\alpha+c_3}}}\frac{d(u^\alpha-c_3)}{u^\alpha}\,\,\ [\mbox{where}\,\,\xi^\alpha+c_3=u^\alpha]\nonumber\\
&=\int_{\sqrt[\alpha]{c_3}}^{{\sqrt[\alpha]{t^\alpha+c_3}}}\frac{du^\alpha}{u^\alpha}\nonumber\\
&=\int_{0}^{{\sqrt[\alpha]{t^\alpha+c_3}}}\frac{du^\alpha}{u^\alpha}-\int_{0}^{{\sqrt[\alpha]{c_3}}}\frac{du^\alpha}{u^\alpha}\nonumber\\
&=(1-\alpha)!\Big[Ln_\alpha({\sqrt[\alpha]{t^\alpha+c_3}})-Ln_\alpha(\sqrt[\alpha]{c_3})\Big]\nonumber
\end{align}
Using $Ln_\alpha x^y=y^\alpha Ln_\alpha x$ we write
\begin{eqnarray}
\int_0^t\frac{d\xi^\alpha}{\xi^\alpha+c_3}=(1-\alpha)!\Big[Ln_\alpha({\sqrt[\alpha]{t^\alpha+c_3}})-Ln_\alpha(\sqrt[\alpha]{c_3})\Big]\nonumber\\=\frac{(1-\alpha)!}{\alpha^\alpha}[Ln_\alpha(t^\alpha+c_3)-Ln_\alpha c_3]\,.
\end{eqnarray}
Similarly 
\begin{eqnarray}
\int_0^t\frac{d\xi^\alpha}{\xi^\alpha-c_3}=\frac{(1-\alpha)!}{\alpha^\alpha}[Ln_\alpha(t^\alpha-c_3)-Ln_\alpha c_3]\,.
\end{eqnarray}
Hence From Eq.(A2-6) we can get
\begin{align}
Ln_\alpha\Big(\frac{z(t)}{C}\Big)&=-\lambda_1\frac{(1-\alpha)!}{\alpha^\alpha}Ln_\alpha(t^\alpha+c_3)-\lambda_2\frac{(1-\alpha)!}{\alpha^\alpha}Ln_\alpha(t^\alpha-c_3)+c_4 \nonumber\\
Ln_\alpha\Big(\frac{z(t)}{C}\Big)&=-\Big(\frac{(1-\alpha)!}{\alpha^\alpha}\Big)[\lambda_1Ln_\alpha(t^\alpha+c_3)+\lambda_2Ln_\alpha(t^\alpha-c_3)]+c_4\,,
\end{align}
where $c_4=-\frac{(1-\alpha)!}{\alpha^\alpha}(\lambda_1+\lambda_2)Ln_\alpha(c_3)$. We have these identities in Jumarie sense\\ $Ln_\alpha(x^y)=y^\alpha Ln_{\alpha}x$ and $(Ln_\alpha(uv))^{\frac{1}{\alpha}}=(Ln_\alpha u)^{\frac{1}{\alpha}}+(Ln_\alpha v)^{\frac{1}{\alpha}}$\\ Here we can make an approximation for $\alpha\approx 1.00$ as\\
$Ln_\alpha(x^y)\approx yLn_{\alpha}x$ and $(Ln_\alpha(uv))\approx(Ln_\alpha u)+(Ln_\alpha v)$. Using these approximations  we write the Eq.(A2-9) 
\begin{align}
Ln_\alpha\Big(\frac{z(t)}{C}\Big)&=-\Big(\frac{(1-\alpha)!}{\alpha^\alpha}\Big)[\lambda_1Ln_\alpha(t^\alpha+c_3)+\lambda_2Ln_\alpha(t^\alpha-c_3)]+c_4\nonumber\\
&\approx -[Ln_\alpha((t^\alpha+c_3)^{\lambda_1})+Ln_\alpha((t^\alpha-c_3)^{\lambda_2})+ln_\alpha c_5]\,\,; c_5=Ln_\alpha c_4\nonumber\\
&\approx -Ln_\alpha[c_5((t^\alpha+c_3)^{\lambda_1})((t^\alpha-c_3)^{\lambda_2})]\nonumber
\end{align}
Hence from above we write the approximate solution for $\alpha\approx 1.00$ as following
\begin{eqnarray}
z(t)=C^{'}(t^\alpha+c_3)^{\lambda_1}(t^\alpha-c_3)^{\lambda_2}\,,
\end{eqnarray}
where $C^{'}=-Cc_5$ is any arbitrary suitable constant. In using the Laplace technique we assume $z(t)\sim t^\delta$ i.e `the power law function' so that is our first approximation and we applied the Laplace relation to get the equation A2-4. The solution yields in terms of Mittag-Leffler function is also sum of power law function. Thus our first approximation is valid to get the final solution in terms of fractional Logarithmic function. Finally too after further simplification and using $\alpha\approx 1.00$ we do get a solution A2-10 i nterms of power law functions. Now for calculation of $\tau$ we chose arbitrarily a value of $\delta=-0.5$ i.e a mid value of $-1<\delta<0$ for different values of $\alpha$. The any other number would have given different values but the analysis would have been similar.
\end{document}